\documentclass[pra,aps,twocolumn,longbibliography,showpacs,floatfix,nofootinbib,superscriptaddress]{revtex4-1}
\usepackage[pdftex]{graphicx}% Include figure files
\usepackage{dcolumn}% Align table columns on decimal point
\usepackage{bm}% bold math
\usepackage{color}
\usepackage{amsthm,amsmath,amssymb,mathtools}
\usepackage{epstopdf}
\def\op#1{#1}

\def\ket#1{| #1 \rangle}

\def\D{\mathcal{D}}

\def\up{{\uparrow}}
\def\down{{\downarrow}}
\def\sx{\op{\sigma}_x}

\def\ss{\rm ss}

\newcommand{\unity}{\leavevmode\hbox{\small1\kern-3.3pt\normalsize1}}
\newcommand{\diss}[1]{\mathcal{D}[#1]}

\newcommand{\beq}{\begin{equation}}
\newcommand{\eeq}{\end{equation}}
\newcommand{\bqa}{\begin{eqnarray}}
\newcommand{\eqa}{\end{eqnarray}}

\newcommand{\dg}{^\dagger}

\newcommand{\BQIC}{Berkeley Center for Quantum Information and Computation}
\newcommand{\DeptChem}{Department of Chemistry, University of California, Berkeley, California 94720 USA}

\newcommand{\berkeley}{\BQIC~and \DeptChem}
\newcommand{\LBNL}{Lawrence Berkeley National Laboratory, Berkeley, California, 94720 USA}

\begin{document}
\title{Backaction driven, robust, steady-state long-distance qubit entanglement over lossy channels}

\author{Felix Motzoi}
\affiliation{\berkeley}
%\footnote{$^*$Current affiliation: Theoretical Physics, Saarland University, 66123 Saarbr\"ucken, Germany}
\affiliation{Theoretical Physics, Saarland University, 66123 Saarbr\"ucken, Germany}

\author{Eli Halperin}
\affiliation{\berkeley}
\affiliation{\LBNL}

\author{Xiaoting Wang}
\affiliation{Hearne Institute for Theoretical Physics,
Department of Physics and Astronomy, Louisiana State University, Baton
Rouge, LA 70803, USA}

\author{K. Birgitta Whaley}
\affiliation{\berkeley}

\author{Sophie  Schirmer} \affiliation{College of Science (Physics),
  Swansea University, Singleton Park, Swansea, SA2 8PP, United
  Kingdom}

%% based on Jan 17 version
\date{\today}

\begin{abstract}
We present a scheme for generating robust and persistent entanglement between qubits that do not interact and that are separated by a long and lossy transmission channel, using Markovian reservoir engineering.  The proposal uses only the correlated decay into the common channel of remotely separated, driven single-photon qubit transitions. This simple scheme is generic and applicable to various experimental implementations, including circuit and cavity QED, with little experimental overhead compared with methods requiring dynamic control, initialization, measurement, or feedback. In addition to avoiding these inefficiencies, the simple protocol is highly robust against noise, miscalibration, and loss in the channel.  We find high quality solutions over a wide range of parameters and show that the optimal strategy reflects a transition from ballistic to diffusive photon transmission, going from symmetrically and coherently driving a common steady state to asymmetrically absorbing photons that are emitted from one qubit by the second.  Detailed analysis of the role of the transmission channel shows that allowing bi-directional decay drastically increases indistinguishability and thereby quadratically suppresses infidelity.  
\end{abstract}

\maketitle

 Deterministically generating remote steady-state entanglement is of fundamental interest for ongoing developments of quantum technologies.  Applications include quantum cryptography, quantum networks, entanglement distillation, scalable quantum computation, and distributed quantum computing~\cite{pielawa2007generation,stannigel2012driven,rafiee2012stationary,joshi2012entanglement,manzano2013synchronization,zippilli2013entanglement}. Much akin to how operational amplifiers have removed many of the timing, calibration, and variability issues in classical circuit technology, offering stabilised entanglement on-demand can serve a similar purpose for quantum technologies, alleviating the need for complex and often inefficient measurement, initialization, photon creation, photon collection, or travel-time synchronization processes.  In this work we propose a scheme for on-demand deterministic generation of remote entanglement that employs reservoir engineering to autonomously arrive at a high-fidelity entangled steady-state solution of qubits in distinct cavities.  Perturbation away from the desired steady-state is self-healing due to the nonlocal relaxation back-action, and therefore naturally robust against noise in ways that pulse and measurement-based generation of entanglement cannot be.

A number of theoretical schemes have been proposed for realizing deterministic steady-state entanglement over short distances, e.g., within a single cavity~\cite{wang2010generating,kastoryano2011dissipative,reiter2012driving,reiter2013steady,Rao2014} or between spatially separated qubits assuming zero or minimal losses in communication~\cite{clark2003unconditional,li2012dissipative,Aron2014steadystate,pichler2015quantum}, and several recent experimental demonstrations have realized above threshold steady-state entanglement, although with limited fidelities~\cite{lin2013dissipative,shankar2013autonomously,Schwartz2015}.  Medium-distance entanglement has also been studied theoretically~\cite{parkins2006unconditional,muschik2011dissipatively,gonzalez2011entanglement,motzoi15} and realized experimentally in~\cite{krauter2011entanglement,Nolleke2013,roch2014observation, Hofmann12heralded}.  However, achieving even postselected, transient entanglement over truly long distances is considerably more challenging, due to the effects of losses.  To date long-distance entanglement has only been realized transiently with very low postselection probability with the use of measurements to postselect entangled states of electron spins~\cite{Hensen2015}.  

We propose an alternative to generate steady-state entanglement between remotely separated qubits.  Our approach relies on generating a contractive map for distant objects using Markovian reservoir engineering~\cite{Schirmer2010} in the presence of a (highly) lossy transmission channel.  The convergence to an entangled steady-state is obtained via the construction of a rank-one decoherence-free subspace~\cite{lidar1998decoherence,DFS} (DFS) and the indistinguishability of the qubits with respect to the destructive environmental interaction. Thus, in the absence of any decay from the two spatially separated objects, this weak (unobserved) measurement projects continuously onto the DFS, essentially constituting an interaction-free measurement~\cite{elitzur1993quantum,vaidman2003meaning,kwiat1995interaction} of the nonlocal state, while the presence of transient leaked information destroys population from any other states, rendering the desired state globally attractive~\cite{Schirmer2010}.  When imperfections in the system are included, this DFS is smoothly varied as a function of the fraction of additional lost information (e.g., in the communication channel), but the coherence of the entangled state can be retained to greatest order by counteracting the dissipative dynamics with constant local unitary drives. We constructively show that these local unitaries in combination with the nonlocal backaction of the environmental interaction are sufficient for global attraction towards a nonlocal state and error correction of perturbation away from this state even in the presence of overwhelming loss.

The current proposal differs from previous work in three respects.  (I) While previous proposals have focused on sidebands and/or two-mode squeezing  \cite{clark2003unconditional,li2012dissipative, lin2013dissipative,parkins2006unconditional,muschik2011dissipatively,krauter2011entanglement} to generate entanglement, we use only two-level systems with single-photon transitions. This reduces the number of decay channels for the qubits and in this respect is more similar to probabilistic measurment schemes which have a single (here unobserved) measurement channel. Reducing the number of drives also simplifies experimental overhead, calibration, and obviates the need for phase-matching between lasers.  (II)  We study the transmission channel between the qubits and its direct effect on the indistinguishability of the two qubits.  We show that not enforcing directionality in the decay channel drastically increases indistinguishability in the low-loss regime, thereby quadratically suppressing infidelity.  (III) We show that while channel loss can exponentially reduce fidelity of protocols that work ideally under low-loss conditions, many other solutions exist that incorporate these additional dissipation effects. These solutions fall into two broad categories corresponding to two regimes for the scattered outgoing photons, namely ballistic (many cavity reflections) or diffusive, with solution parameters varying from symmetric to highly asymmetric along the transition.

\begin{figure}
\includegraphics[width=0.9\columnwidth]{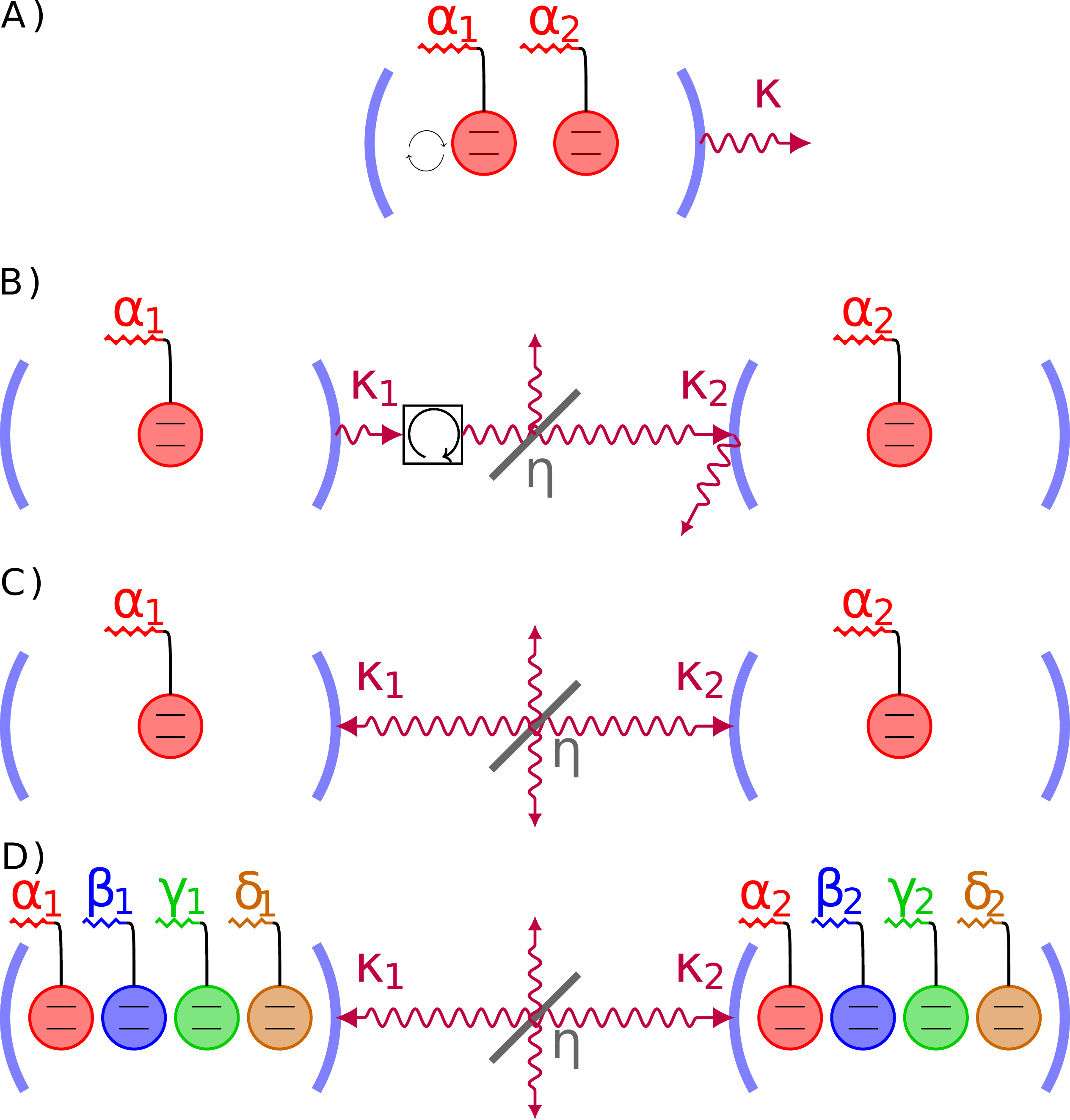}
\caption{(Color online.) (A) Stabilisation of two qubits coupled to a single cavity undergoing correlated decay. (B) Cascaded system setup: the qubits are placed in cavities separated by a transmission line and circulator element such that decay is in one direction.  (C) Bidirectional cascaded  decay, i.e.~without a circulator. (D)~Multiple qubit setup, such that qubits become pairwise entangled in the two cavities. } \label{fig:setup}
\end{figure}

The experimental protocol we present below is inherently capable of deterministically and persistently generating long-range above-threshold entanglement while being robust to a large amount of noise or miscalibration in the system parameters. It consists of a nonlocal, weak, destructive measurement interaction (which remains unobserved), and always-on, constant local rotation operator.   The destructive measurement here corresponds to a correlated atomic decay operator~\cite{dicke1954coherence}, which has multiple dark states, including a Bell state and the ground state. The role of the rotation operators is to remove the degeneracy in the dark states by acting trivially only on the desired Bell state, while rotating the other states amongst themselves.  Because the dark state is unique, all initial states must converge asymptotically onto this state~\cite{Schirmer2010}.  

Section \ref{sec:loss-less} introduces our mechanism and formalism to generate nonlocal states.  Applications are given for a single-cavity setup (Section \ref{sec:single}), cascaded cavities (Section \ref{sec:cascade}), and multiple reflections off both cavities (Section \ref{sec:bidir}).  The scheme is applicable to both microwave and optical domain cavity quantum electrodynamics, using atomic or electrical (superconducting) qubits.  Section \ref{sec:robustness} discusses in detail the robustness of the protocol, while Section \ref{sec:optim} discusses optimality of the protocol and scalability. We conclude and give an outlook in Section \ref{sec:conc}.
 
\section{loss-less solution}
\label{sec:loss-less}

The correlated decay mechanism corresponds to a common output channel of the atomic decay operators and is given by the Lindblad master equation
\begin{subequations}  \label{mastereq}
\begin{align}
   \dot\rho &=-i \sum_j [H_j,\rho]+\sum_k\D[L_k]\rho \\
   L_1&=s_{1}\sigma_1^-+s_{2}\sigma_2^-
\end{align}
\end{subequations}
where the dissipation is augmented by local dynamics
\begin{equation}
H_j=\alpha_j\sigma_j^++\alpha_j^*\sigma_j^-+\Delta_j \sigma_j^+\sigma_j^-
\end{equation}
where $\sigma^-_j$ is the lowering operator of the $j$-th qubit and $s_j$, $\alpha_j$ are its decay rate and Rabi frequency, respectively.  We work in a rotating frame such that the $j$-th qubit is detuned from the drive by $\Delta_j=\omega_j-\omega_d$. The common drive frequency $\omega_d$ for both qubits is an important requirement for the operation of the protocol.  This class of master equations has a rich space of solutions, a large subset of which feature steady-state entanglement.  In what follows we consider three different architectures that reproduce the same master equation formulation without loss, as well as their prevalent loss mechanisms and their mitigation.

For the particular solution, $\alpha_j=\alpha$, $\Delta_1=-\Delta_2=\Delta$ and $s_j=s$, it is easy to verify that
\begin{equation}
 \ket{\Psi_{\ss}}= \Delta \ket{\up\up} + \alpha\ket{\up\down} -\alpha \ket{\down\up}
\label{eqpuresteady}
\end{equation}
is an eigenstate with eigenvalue $0$ of the effective Hamiltonian $H$ and the Lindblad operator $L_1$, and thus a steady state of the dynamics~\eqref{mastereq}. For $\Delta=0$ (no detuning),  the system is decomposable and there are infinitely many other steady states, rendering $\ket{\psi_{\ss}}$ non-attractive~\cite{Schirmer2010} when starting from any other state. However, increasing $\Delta$ infinitesimally will break the symmetry between the steady states, rendering $\ket{\psi_{\ss}}$ the \emph{unique globally attractive} steady state~\cite{Schirmer2010}.  The concurrence of this steady state is $C(\Psi_{\ss})=2\alpha^2/(\Delta^2+2\alpha^2)$, hence $\Delta/\alpha\approx 0.1$ gives $C(\Psi_{\ss})=0.999$.  While the given solution has the simplest form, it is by no means the only regime where entanglement can be generated.  For any values of qubit decay rates $s_1$, $s_2$, and qubit frequencies $\omega_1$, $\omega_2$, there are drive amplitudes and frequencies that will maximize concurrence.  Such solutions over a wide range of parameters can be easily found analytically, numerically, or experimentally.

\section{Single-cavity decay}
\label{sec:single}

As a prelude to analysis of distant qubits in remote cavities, we first consider the most straightforward architecture to obtain the dynamics~\eqref{mastereq}, which is by coupling two (artificial) atoms to a common cavity mode operating in the same frequency range~\cite{majer2007coupling,khudaverdyan2009quantum}.  Dissipative generation of steady-state entanglement in this architecture was already studied under loss-less conditions in~\cite{wang2010generating}. Here we extend the analysis to incorporate the effects of losses due to dephasing and relaxation of the qubits, which will also be an important component of our subsequent analysis for distant qubits.

For simplicity we take the atoms to have individual drive and measurement control lines, although these can be realized through the cavity.  Neglecting again any direct coupling between the qubits/atoms  (which is typically weak and can be suppressed by detuning), the joint atom-cavity dynamics are
\begin{subequations}
   \label{singlecavity} 
  \begin{align}
    H^{JC}_{j}(a)  &= H_j+g_j(\sigma_j^- a\dg + \sigma_j^+ a) \\
    \dot\rho     &= \textstyle -i \sum_j [H^{JC}_j(a)+\delta a\dg a, \rho]+\D[\sqrt{\kappa}a]\rho,
\end{align}
\end{subequations}
where $g_j$ is the interaction strength between atom $j$ and the cavity, and $\delta$ and $\kappa$ are the detuning and decay rate of the cavity, respectively.   Adiabatically eliminating \cite{shore1993jaynes} the resonator mode(s) via the unitary transformation
\begin{equation}
   \label{Tj:single}
  T_j=\exp((s_j a \sigma_j\dg - s_j^* a\dg\sigma_j^-)/\sqrt{\kappa})
\end{equation}
results in Eq.~\eqref{mastereq} with $T_jH_j^{JC}(a)T_j\dg=H_j$ and
\begin{equation}
   s_j=\frac{\sqrt{\kappa}g_j}{\delta-\Delta_j+i\kappa/2}.
\end{equation}

\begin{figure}
  \includegraphics[width=0.8\columnwidth]{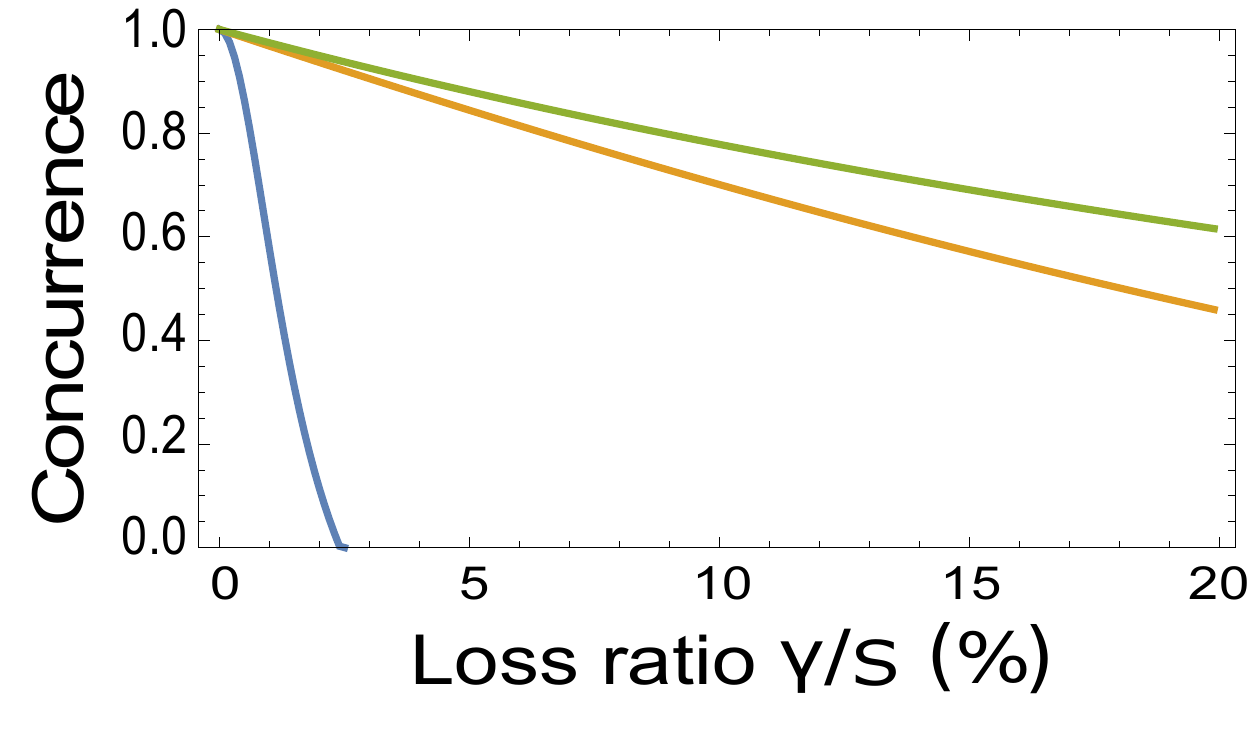}
  \caption{(Color online): Steady state concurrence $C_{\ss}$ as function of the ratio of intrinsic qubit decay $\gamma$ to cavity mediated decay $s$. The bottom blue line is when the simple solution \eqref{eqpuresteady} is used with $\Delta/\alpha=0.01$ and $\gamma=\gamma^r_1$.  The middle orange line is for $\gamma=\gamma^r_1$ and the analytical first order solution $\Delta/\alpha=\sqrt{l/2}$. The top green line is for $\gamma=\gamma^\phi_1$ and the analytical first order solution $\Delta/\alpha=\sqrt{l/2}$.}
  \label{figCvsLoss1Cav}
\end{figure}

In addition to the cavity-mediated dynamics, the (artificial) atoms typically have other intrinsic dissipation mechanisms, such as local relaxation and dephasing given by the Lindblad operators
\begin{subequations} \label{eqT1T2}
\begin{gather}
  L_2         =  \gamma^r_1\sigma_1^-, \qquad
  L_3         = \gamma^\phi_1\sigma_1^z,\\
  L_4         =  \gamma^r_2\sigma_2^-, \qquad
  L_5         = \gamma^\phi_2\sigma_2^z.
  \end{gather}
\end{subequations}
The lasting effect of these processes is to destroy the purity of the stabilised state. For a constant ratio $\Delta/\alpha= 0.01$ of Rabi to drive detuning frequency, the lower blue line in Fig.~\ref{figCvsLoss1Cav} demonstrates the drastic falloff in concurrence as a function of unwanted environmental couplings, here $\gamma=\gamma^r_1$. This provides further evidence that intrinsic single qubit relaxation processes can greatly reduce expected fidelities of entangled states created using engineered correlated decay between the qubits.

One way to mitigate such detrimental effects is by proportionally increasing the strength of the dominant (desired) dynamics by increasing the ratio $s_j/\gamma_j$, e.g. by increasing the transmissivity of the output port of the cavity or the qubit-cavity coupling.  In other words, increasing the correlated decay moves the dynamics towards the left side of the graph.  However, physical constraints impose bounds on achievable $s_j$.

A complementary and more practical solution is optimizing the dynamics to take into account the form of these deleterious effects.  While the intrinsic loss operators act primarily by decreasing the purity of the entangled state, these rates can be reduced to greatest order by making dissipation out of the (approximately pure) steady-state more energetically unfavourable, that is, by essentially causing destructive interference in the transient population outside of the subspace. Of course, as seen from the approximate solution \eqref{eqpuresteady}, increasing the detuning between qubits (relative to Rabi frequency) will also incorporate more of the ground state into the stead-state superposition, and so there is a trade-off between increasing purity and decreasing the ground-state component.

The general form of the steady state for the two qubit system can be determined by solving the system of $d^2-1$=$15$ differential equations for the elements of the stabilised density matrix or equivalent Bloch vector.  This solution will be an 8th order polynomial whose concurrence~\cite{Wooters1998} can be maximized.  Using the first order perturbation in the small parameter $l=\gamma/s$ gives optimal choice $\Delta/\alpha=\sqrt{l/2}$.  This solution is plotted in Fig.~\ref{figCvsLoss1Cav} in green (top line) for $\gamma=\gamma^r_1$ and orange (middle line) for $\gamma=\gamma^\phi_1$. It is evident that both solutions show pronounced resilience to loss.

Tuning of this ratio is straightforward by modifying the Rabi frequency, but alternatively the qubit frequencies can also be tuned via static magnetic fields in both cavity and circuit QED. This also allows for indirect tuning of the amplitude and phase of $s_j$.  Modifying the phase between $s_1$ and $s_2$ enables preparation of different target Bell states, while modifying the amplitude controls the entanglement convergence rate. Operating the qubits close to the cavity and drive frequencies allows for generating entanglement at the fastest rate while larger detuning minimizes decay.

Moving away from symmetric solutions, it is straightforward to find similarly robust solutions. For example, letting $s_1=0.8s_2$ and $\gamma^r_1=0.02s_1$, we optimize the local drives to values of  $\alpha_1=0.88s_1$, $\alpha_2=0.79s_1$, $\Delta_1=0.28s_1$, $\Delta_2=0.48s_1$, allowing for 95\% concurrence. Thus, even if the qubits are fixed or parked at their optimal working points to minimize intrinsic decoherence  such asymmetry is not an impediment to achieving high grade concurrence.

\begin{figure}
\includegraphics[width=0.80\columnwidth,height=1.6in]{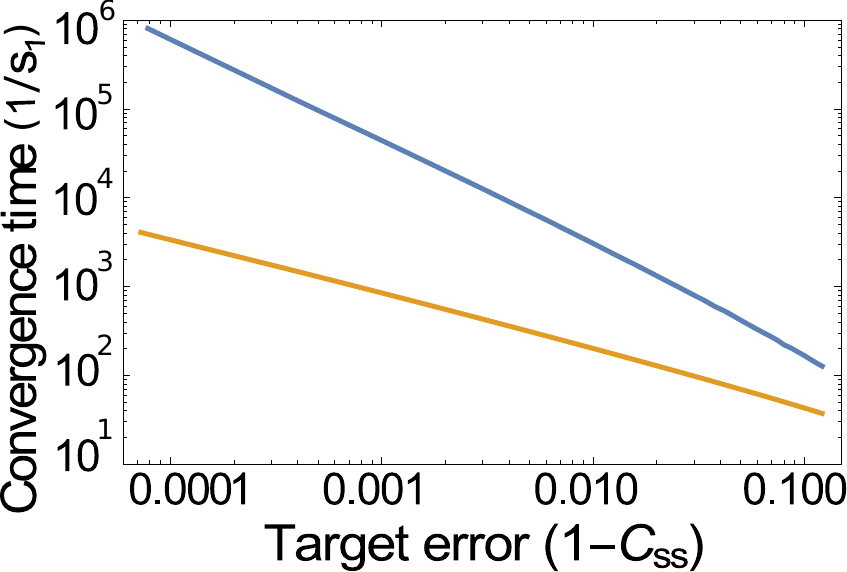}
\caption{(Color online) Convergence time as a function of achievable concurrence $C_{ss}$ for respective intrinsic loss in a single cavity setup for constant detuning to Rabi frequency ratio (blue) and exponentially decreasing ratio with constant offset, as per Eq.~\ref{eq_expdetuning} (orange).} 
  \label{fig3}
\end{figure}

Meanwhile, the rate of convergence to the steady state is determined by the real parts of the eigenvalues of the Liouvillian, specifically the eigenvalue with the smallest non-zero negative real part.  We investigated the time required to reach the steady-state through numerical simulation of the dynamics~\eqref{mastereq} with \eqref{Tj:single}.  This time was found to be virtually independent of the initial separation of the states.  For solutions with finite detunings, the convergence time is very fast, on the order of $s$.  However, the convergence time increases approximately linearly with $\alpha/\Delta$ as the eigenvalues become increasingly closer, as is plotted by the top blue line in Fig.~\ref{fig3} for near-optimal $s_j=2\alpha$.  Nonetheless, even for very small detunings, these rates compare favourably to entanglement generation rates involving combinations of photon pair creation, detection, heralding, collection, or other mechanisms with much smaller than unity efficiencies.  Furthermore, the rate of convergence using dissipation can be significantly accelerated by starting with a large detuning (relative to the coupling strength) and exponentially reducing it to the optimal value, via
\begin{equation}
  \label{eq_expdetuning}
  \Delta(t)=\alpha\sqrt{l/2}+e^{-\alpha\sqrt{l/2} t}.
\end{equation}
This solution is plotted as the bottom orange line and shows order(s) of magnitude improvement in the convergence time relative to the static case.

The entangled state is globally attractive and thus intrinsically robust with regard to perturbations, producing persistent entanglement with minimal experimental overhead, removing timing issues and inefficiencies.  Related protocols for short-distance, lossless entanglement have been developed in~\cite{kastoryano2011dissipative,li2012dissipative}.  The next two sections present the main results of this paper, which derive from extending this analysis to qubits separated by large distances with associated large distance-dependent losses.

\section{Cascaded decay}
\label{sec:cascade}

The model \eqref{mastereq} is still useful if the qubits to be entangled are placed in separate cavities connected by a transmission line instead of the same cavity.  In this case the dynamics can be modeled using the cascaded systems theory~\cite{gardiner1993driving,carmichael1993quantum} or modern SLH quantum network theory~\cite{Gou.Jam-2009,Gough:2012fl}.  Both produce the same Markovian dynamics but we utilize the latter here for compactness.  Each cavity is described by Eq.~\eqref{singlecavity} with the scattering/Lindbladian/Hamiltonian triplet
\begin{equation}
  \{ S_j,L^c_j,H_j^c\} = \left\{-\emph{1},
  \begin{bsmallmatrix}
    \sqrt{\kappa_j}a_j \\0
  \end{bsmallmatrix} ,
    H^{JC}_j(a_j)+\delta_j  a_j\dg a_j\right\}
  \end{equation}
for the $j$-th (single-mode) cavity with annihilation operator $a_j$, detuning $\delta_j$ and decay rate $\kappa_j$. Flow along the transmission line is given by the scattering matrix $S^t$, which effectively relays the output field of one cavity to the input of the other with efficiency $\eta$, with the rest of the field being lost along the way, and
%BW 1208
can be modeled as a beam splitter, $S^t = \eta I + i\sqrt{1-\eta^2}\sx$.  Loss in such a long-distance architecture is much more troublesome than in the single cavity case because the single-qubit loss and correlated-decay are proportional, and we therefore cannot simply increase one over the other.  Furthermore, loss is expected to grow exponentially with distance and is a primary impediment to remote entanglement proposals.

For such unidirectional transmission the effective dynamics are compactly obtained from the SLH composition formulae for cascaded systems,
\begin{subequations}\label{eqcascadedcavities}
  \begin{align}
  \begin{bsmallmatrix} L_1 \\ L_2\end{bsmallmatrix}
   &= L_2^c - S^t L_1^c
      = \begin{bsmallmatrix}
          \eta \sqrt{\kappa_1}a_1-\sqrt{\kappa_2}a_2 \\
          \sqrt{\kappa_1(1-\eta^2)}a_1
          \end{bsmallmatrix}\\
   H^c_{12}
    &= {L_2^c }\dg S^t L_1^c+\mathrm{h.c.}
      =i\eta\sqrt{\kappa_1\kappa_2}(a_1\dg a_2-a_2\dg a_1),
  \end{align}
\end{subequations}
such that $\dot\rho=-i[H^c_{12}+\sum_j H_j,\rho]+\sum_j \D[L_j]\rho$.  In practice, directionality is enforced by placing a circulator in the optical path to prevent (i.e.~destroy) back-reflection from the second cavity, similarly to what is done for measurement-based approaches~\cite{kerckhoff2009physical, roch2014observation, motzoi15}.  This can also be seen from looking at the Heisenberg equations of motion for Eq.~\eqref{eqcascadedcavities} and checking that the only the second cavity has dependence on the state of the first but not vice-versa. Directionality also ensures that relaxation of the first qubit cannot result from direct interaction with the second and that therefore the relaxation operator's backaction must be nonlocal, as well as any error-correction that results, exhibiting different behaviour to conventional (i.e. Shor-type) error correction where feedback of the syndrome information is needed~\cite{ShorErrcorr}.

In this effective frame, we now perform component-wise adiabatic elimination \cite{gough2010} of the cavity degrees of freedom, as was done for Eq.~\eqref{singlecavity}, using now
\begin{equation}
  \label{eqdecayamplitude}
  T_j    = \exp\left[(s_{j}a_j \sigma_j \dg - s_{j}^* a_j\dg \sigma_j)/\sqrt{\kappa_j}\right],\quad
  s_{j} = \sqrt{\kappa_j}{g_j}/{\tilde\Delta_{j}},
\end{equation}
where $\tilde\Delta_{j}$=$\delta_j$-$\Delta_j$+$i\kappa_j/2$.   We then obtain the reduced qubit master equation
\begin{subequations}
  \begin{gather}
  \dot\rho  = \textstyle -i \left[H_{12}+\sum_j H_j,\rho\right] +\sum_j \D[L_j]\rho \\
  L_1         = \eta s_{1}\sigma_1^-+s_{2}\sigma_2^-, \qquad
  L_2         = \sqrt{1-\eta^2}s_{1}\sigma_1^-.\\
   H_{12}  =  i\eta(s_1^* s_2 a_1\dg a_2- s_2^* s_1 a_2\dg a_1)
  \end{gather}
\end{subequations}
These dynamics reduce to Eq.~\eqref{mastereq} in the limit of perfect transmission, $\eta\to 1$.  Otherwise, the loss channel takes the operator form of relaxation on the first qubit, just like $L_2$ in the single cavity case \eqref{eqT1T2}, but with loss $l=\sqrt{1-\eta^2}$ now dependent on the channel inefficiency (i.e.~the distance travelled). In these detrimental conditions, the steady-state of the system becomes again very quickly mixed, containing all four basis states.  However, the general form of the steady state can once again be solved analytically or numerically for this new master equation, and subsequently optimized. In the low loss limit, the maximum concurrence can be achieved for e.g.~$\Delta_j=0$, $\alpha_j=1$, $s_2=\frac{1}{5}+2\sqrt{l}+l$, $s_1=s_2+8l^2$.  Note that no detuning is needed here because the effective inter-qubit coupling naturally splits the energies of levels in the single-excitation subspace. Many other optimal and near-optimal solutions exist to the steady state, e.g. for finite detuning $\Delta_j$.

\begin{figure}
  \includegraphics[width=0.9\columnwidth]{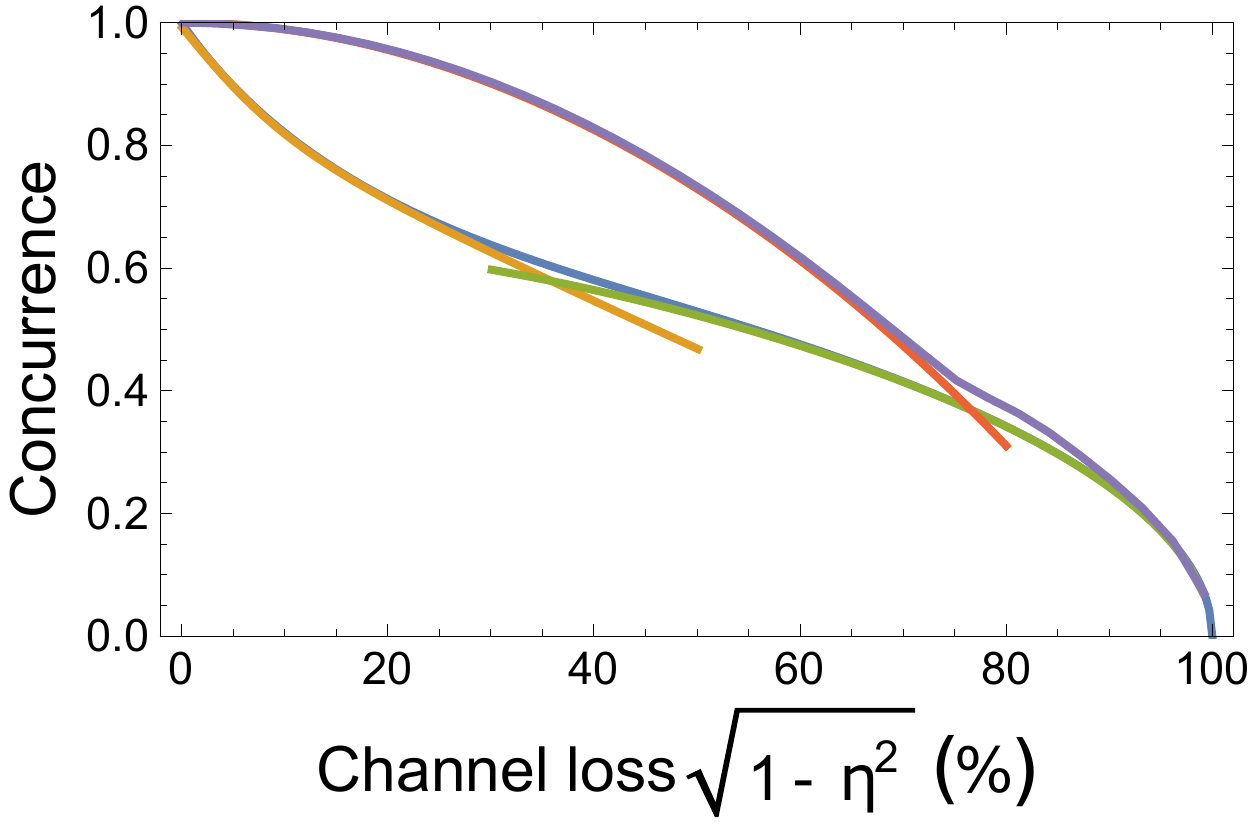}
  \caption{(Color online): Steady state concurrence $C_{\ss}$ as function of channel inefficiency for qubits in remote cavities.  The bottom three lines are for cascaded decay (Eq.\ref{eqcascadedcavities}) and the top three lines are for bi-directional decay as in  Eq.~\eqref{eqbidirectional}. Analytical low and high-loss approximations are plotted on top of the numerical optimization results.}
  \label{figCvsLoss}
\end{figure}

In the high loss limit $\eta\to0$, we obtain a solution $\Delta_j=0$, $\alpha_1=3/4$, $\alpha_2=1/4-\eta$, $s_1=1$,  $s_2=5-2\sqrt{\eta}$. This high-loss solution differs from those in the low loss limit in that it is highly asymmetric between the two qubits. In essence, we find that the  extreme loss out of the first qubit can be counteracted by decreasing the cavity decay and increasing the Rabi strength, such that lost population is pumped back into the excited state.  Meanwhile the second cavity is  strongly coupled and weakly pumped, such that at least some of the population from the first qubit finds its way into the second qubit and stays there (specifically, whenever no field leaks out of the transmission channel). This ensures that the dark-state superposition will contain at least some coherent population in the single-excitation subspace.  The concurrence of the low-loss and high-loss solutions as a function of the inter-cavity loss are plotted in Fig.~\ref{figCvsLoss} as the lower yellow and green lines respectively, together with numerical optimization results (lower green curve).  The results show optimality of the simple analytic solutions over a wide regime and resilience under even overwhelming loss.

\section{Bidirectional decay}
\label{sec:bidir}

Whereas including a circulator retains the important properties of our solution, such as persistence, determinism, and robustness against losses, there are at least two reasons for not including it.  The first is that the circulator element will be imperfect and decrease the transmissivity of the channel, although  efforts exist to create new designs that minimize such losses~\cite{metelmann2015nonreciprocal}. The second is that allowing for bi-directional transmission means that a photon will travel on average $\bar n$ times through the waveguide and therefore it will become harder to distinguish  from which qubit the information that leaks out is coming from. This means that the mixing/ground state contribution in the steady state can in principle be less pronounced.  

When the transmission channel is allowed to flow in both directions, the system can be modeled as two unidirectional systems forming a loop together.  The internal scattering edges from one cavity to the other in the SLH representation correspond to beam-splitter elements $S^t(1,1)$ while output edges correspond to $S^t(1,2)$. Thus,
\begin{subequations}
  \begin{align}
  S_{e,i} &= \begin{psmallmatrix} \sqrt{1-\eta^2} \\0\\ \sqrt{1-\eta^2}\\0 \end{psmallmatrix}, \;
  S_{i,i} = \begin{pmatrix} \eta &0&0&0 \\  0&-1&0&0\\0 &0 & \eta&0 \\  0&0&0&-1\end{pmatrix}, \; \\
    A      &= \begin{pmatrix} 0 & 0 & 0 & 1 \\ 1 &0 & 0 &  0\\ 0 &1 & 0 &  0\\ 0 &0 & 1 &  0 \end{pmatrix}
  \end{align}
\end{subequations}
where $A$ is the adjacency matrix for internal edges taking the output of each cavity to a beamsplitter and into the other cavity, with indices $i$ and $e$ corresponding to internal and external edges, respectively.  $\eta$ can in general be taken as complex ($\eta=|\eta|e^{i\phi}$) to take into account the phase shift for the given path length in addition to loss through the fiber  (see also \cite{pichler2015quantum}). Computing the linear fractional transformation for elimination of the internal edges in the setup results, in the Markovian limit, in
\begin{subequations}\label{eqbidirectional}
  \begin{align}
  \begin{bsmallmatrix} L_1 \\ L_2 \end{bsmallmatrix}
   &= {S_{e,i}}\dg  \left(S_{i,i}-A\right)^{-1}
         \begin{bsmallmatrix}
         \sqrt{\kappa_1}a_1 \\ 0\\\sqrt{\kappa_2}a_2 \\0
         \end{bsmallmatrix}
   =   \begin{bsmallmatrix}
            \eta \sqrt{\kappa_1}a_1+\sqrt{\kappa_2}a_2 \\
            \sqrt{\kappa_1}a_1+\eta\sqrt{\kappa_2}a_2
         \end{bsmallmatrix} \\
   H&= \sum_j H^{JC}_{j}(a_j)+\mathrm{Im}(\eta)\sqrt{\kappa_1\kappa_2}(\eta a_1\dg a_2+a_2\dg a_1),
  \end{align}
\end{subequations}
such that $\dot\rho=-i[H,\rho]+\sum_j\diss{L_j}$, after normalization. It is straightforward to see that the decay operators $L_1$ and $L_2$ here are less susceptible to loss than those in Eq.~\eqref{eqcascadedcavities}.  For instance, setting $\eta \sqrt{\kappa_1}=\sqrt{\kappa_2}$ and using $l=\sqrt{1-\eta^2}$, we see that the asymmetry in $L_2$ will be proportional to $l^2$ inside the dissipator, while for the cascaded setup it will linearly proportional to the loss $l$.

Given the master equation, we can once again perform component-wise adiabatic elimination, as for Eq.~\eqref{singlecavity} and 
%BW 1208
thereby arrive at the reduced qubit equation given by
\begin{subequations}
  \begin{gather}
   \dot\rho = \textstyle -i \sum_j [H_j,\rho]+\sum_j \D[L_j] \rho\\
   L_1        =  {\eta s_{1}\sigma_1^-+s_{2}\sigma_2^-}, \quad
   L_2        = {s_{1}\sigma_1^-+\eta s_{2}\sigma_2^-} 
  \end{gather}
\end{subequations}
with the same parameters as before. The dynamics reduce to Eq.~\eqref{mastereq} in the limit of perfect transmission, $\eta\to 1$.

In the low-loss limit an analytic solution for optimizing the concurrence is attained for $\phi=\pi$, $s_j=s$, $\Delta_1=-\Delta_2=l s$, $\alpha_1=-\alpha_2=1.7s$.  Note the dependence in this case of the parameters on $l$, rather than $\sqrt{l}$ as was the case for the cascaded setup, indicating loss of purity  that happens quadratically slower so that there is now a greater resilience to loss. The concurrences of this solution as a function of channel loss are plotted in red and compared with the previous results from cascaded decay in Fig.~\ref{figCvsLoss}, where indeed additional quadratically better immunity to loss is clearly visible.

In the high loss limit, $\eta\rightarrow0$, the optimal concurrence is attained for e.g.~$\phi=\pi$, $\Delta_j=\alpha_2=0$, $\alpha_1/s_1$=$1-\sqrt{\eta}$, $s_2=1.7s_1$.  Interestingly, in this high loss limit, the bi-directional architecture does not produce quantitatively higher concurrence than the cascaded setup, owing to a similar strategy that does not use the symmetry of the setup. For high loss, it is once again advantageous to choose a highly asymmetric solution where the second cavity is not driven but strongly coupled to the transmission channel so that any photons that do make it to the second cavity consequently excite the second qubit.  The numerical optimization results are also plotted (in purple) in Fig.~\ref{figCvsLoss}, once again showing the near optimality over a wide range of the analytical formulas. Clearly the bi-directional case outperforms the cascaded setup over a wide range.

\section{Robustness and Experimental Feasibility}
\label{sec:robustness}

We now discuss the robustness with respect to a number of aspects of experiment.

The first aspect is global convergence of the protocol regardless of the initial state and final time.  This means that there is no need to precisely calibrate or cool to the initial state to produce the correct final state.  There is  also no need to precisely know how long it will take for a wavepacket to travel or to synchronize arrival times or generation of signals. Moreover, the robustness is also towards transient perturbations away from the steady-state,  since any temporary action on the state will be subsequently undone and the optimal state restabilized.

\begin{figure}
\includegraphics[width=0.95\columnwidth]{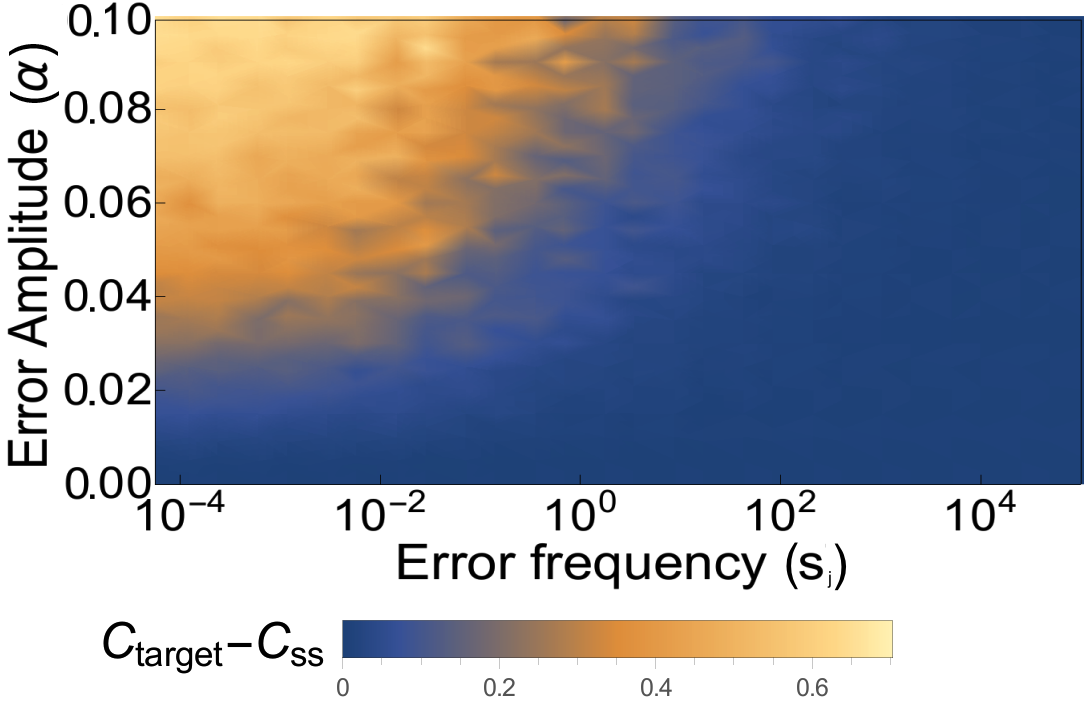}
\caption{(Color online) Drop in concurrence $C_{ss}$ as a function of  deviations (with 10\% maximum) from expected value of Rabi frequencies $\alpha_i$ and mean inverse correlation time of the error.   The deviations are chosen anti-symmetric for the two qubits (worst-case error). The system is simulated for the single-cavity master equation (\ref{mastereq}) with Eq.~\ref{eqT1T2} set to $\gamma^r_1/s=0.15$.}
 \label{figfreq}
\end{figure}

To see this, we can quantify how robust the experiment is against various frequencies and forms of noise sources.  For example, we quantify the effect of extra noise on the control ports of the qubits. We model this as random telegraph noise~\cite{Mottonen2006} on top of the Rabi frequencies which is chosen to be anti-symmetric between the controls to reflect the worst case (the symmetric case is almost immune to noise). Fig.~\ref{figfreq} shows the effect on the steady concurrence of different amplitudes and inverse correlation times of this form of noise.  Indeed we see that for very slow noise there is some dropoff in concurrence compared to the target value of 97\% (the error roughly doubles for a 2\% drift/miscalibration in the Rabi frequency) while there is almost no effect for fast noise.

\begin{figure}
\includegraphics[width=0.95\columnwidth]{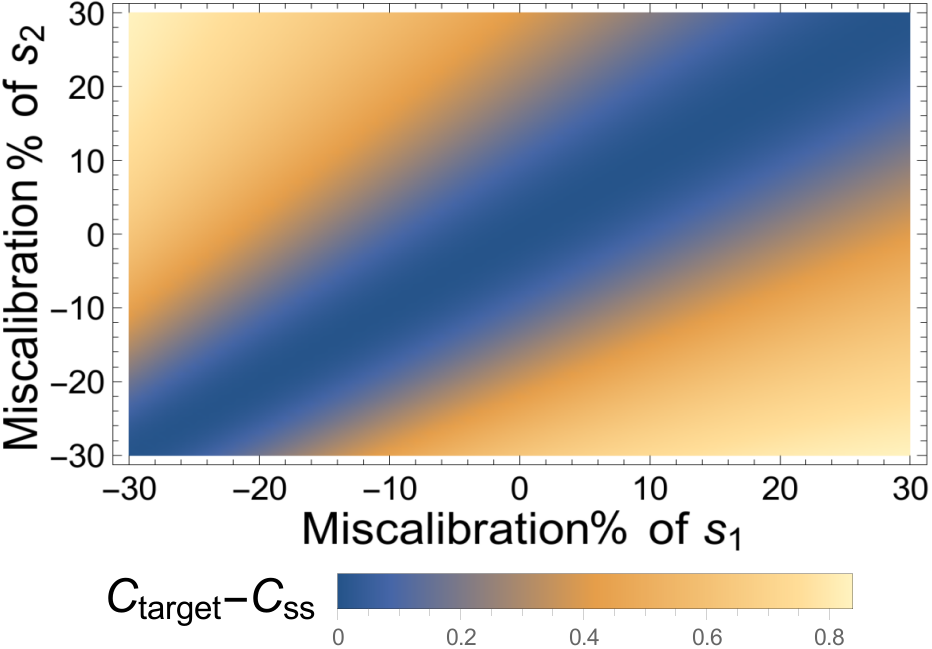}
\caption{(Color online) Drop in concurrence $C_{ss}$ as a function of  deviation (30\% maximum) from expected values of qubit decay rates $s_1$ and $s_2$ for a single cavity, with other parameters the same as in Fig.~\ref{figfreq}.}
  \label{figcalib}
\end{figure}

While there is expected to be little drift in the parameters on the scale of the convergence time, if the stabilization is allowed to run unattended  for long intervals of time, there can be small shifts in certain parameters.  Moreover, certain parameters such as the cavity decay rate can be non-trivial to measure, so small calibration errors may arise.  In Fig.~\ref{figcalib}, we  plot  the concurrence as a function of deviations from the optimal  parameter values for both decay rates of the qubits.  We see that there is indeed robustness against this parameter, with miscalibrations up to around 10\% being tolerable when the drift is anti-symmetric, and almost complete robustness when the shift is symmetric for the two qubits.  Robustness to qubit frequency drift is even greater, with about 20\% deviation being tolerable. 

Another significant robustness of the experiment comes in the context of the design parameters of the experiment.  While the decay rates and detunings of the qubits can be tuned by shifting the qubit frequencies (e.g. by magnetic fluxes or Stark shifts), this is typically not even required.  As discussed in the previous sections, a very wide range or combination of experimental parameters can produce near-optimal concurrence for the losses present, simply by changing the frequency and amplitude of the qubit drives.  For example, for asymmetric decay $s_1=2/3 s_2$ and large 40\% loss, it is still possible to robustly achieve 80\% concurrence with the appropriate driving fields.  Finding the optimal value can be performed straightforwardly either analytically or numerically as above, or during the experiment using simple gradient-free optimization routines such as Nelder Mead~\cite{neldermead} or quasi-Newton methods with finite different gradient estimation~\cite{PhysRevA.80.030301}.

While the concurrence is a preferred figure of merit, we obtain similar fidelities for overlap fidelity of the steady state with the singlet Bell state.  For example, choosing an expected bidirectional waveguide loss of $60\%$, the overlap fidelity with the Bell state is still $71\%$ while concurrence is $61\%$.  Because we have not optimized for this metric, this figures can likely be further improved by sacrificing some coherence for more population inside the odd excitation number manifold.

Although the parameters used are dimensionless and thus work at various timescales, they are also grounded experimentally and are well within the range of typical parameters.  For example, for contemporary circuit-QED architectures~\cite{roch2014observation, Schwartz2015},  cavity decays ($\kappa$) and detunings  ($\Delta$) in the low MHz, qubit-cavity couplings ($g$) and Rabi frequencies  ($\Omega$) on the order of 10-100~MHz are accessible, and would permit convergence times on the order of $1/s\sim {\kappa}\Delta/g$ or about 100~ns, much faster than typical coherence times. Provided the steady-state and parameters are appropriately modified to account for the extra losses (as described in the single cavity case), a Purcell to intrinsic decay ratio $s/\gamma>100$ would affect concurrences by  $<1\%$, with steady-state concurrence corresponding to losses in the transmission line according to Fig.~\ref{figCvsLoss}.

In the optical domain, similar couplings and convergence rates could be achieved with cavity-QED using trapped atoms~\cite{reiserer2014quantum}. Increasing the strengths of vacuum couplings, Rabi frequencies, detunings, and cavity decays to the $\sim1$~GHz range, e.g. coupling NV centres to photonic crystal cavities \cite{yang2011quantum}, would permit convergence within nanoseconds.  Once again, for experiments with $s/\gamma>100$, nearly all intrinsic loss could be counteracted to about $99\%$, while a more conservative estimate of $s/\gamma>10$ would still permit a baseline concurrence of $75\%$, both well within the range needed for e.g. entanglement distillation.

Finally, we also consider how well the protocol performs for large distances between the cavities.  Since photon loss in the channel (measured in dB) will be proportional to the distance traveled, we have optimized the concurrence for various amounts of loss.  For a conservative estimate of 0.1dB loss per meter of microwave coaxial cable~\cite{PhysRevLett.112.170501}, we expect there still to be considerable entanglement when microwave cavities are tens of meters apart.  For optical waveguides loss rates are expected to be even better ($<1$dB/km)~\cite{nagayama2002ultra,roberts2005ultimate} so that steady-state entanglement ought to be achievable across multi-kilometer separations.

\section{Extensions and optimality}
\label{sec:optim}

The protocol we have presented is very simple and so lends itself well to generalization and extension.

One aspect that can be readily generalized is the relative phase of the entangled state.  By changing the relative phase of the qubit decay amplitudes $s_1$ and $s_2$ (and accordingly also the relative phases of the Rabi controls) it is possible to obtain any entangled state in the single-excitation subspace. This can be tuned by changing the relative real and imaginary parts of $s_i$ through the cavity-to-qubit detuning (see e.g. Eq.~\eqref{eqdecayamplitude}) or by introducing phase shifters in the transmission channels.  For bi-directional decay, both of these allow for calibration of the inter-cavity phase $\phi$ as well.

We can also consider whether the protocol can be further generalized to be more robust to transmission loss.  As far as Markovian two-qubit implementations are concerned, the answer appears to be no.  We have optimized over general Lindbladian operators of the form
\begin{equation}
\mathcal{L}\rho= \diss{\kappa_1\hat O_1+\eta\kappa_2\hat O_2}\rho +
\diss{\eta\kappa_1\hat O_1+\kappa_2\hat O_2}\rho
\end{equation}
and no solution was found better than the upper bound given in Fig.~\ref{figCvsLoss}, indicating the relative optimality of our solution, at least for Markovian solutions.  Of course solutions that use auxiliary degrees of freedom (such as in entanglement distillation~\cite{entangdistill1996}) can in principle produce higher concurrences at the cost of a larger Hilbert space.

Including non-Markovian effects is also a direction of future research, for example using the method presented in Ref.~\cite{Grimsmo_time2015}.  This is not necessary in the uni-directional case since the only effect of the transmission channel is to delay its eventual environmental measurement, such that the first and second qubit are correlated instead at different times (eventually reaching a common entangled steady-state).  For bi-directional decay, the state of the qubits may change between different reflections of the light off the cavities, and so the condition of the SLH formalism that $s\ll L/c $ is formally needed; but in practice, short-time transient effects do not pose much issue (as seen in Fig.~\ref{figfreq}) such that the present results are expected to hold even entering into the non-Markovian regime. Moreover, only the convergence rate and not the steady-state itself can be affected by non-Markovianity, since in the steady-state there are no memory effects.

It would also be interesting to consider to what extent the protocol gets around the need for error correction by syndrome measurement and feedback.  For the regimes considered here this has a drastic effect on the ultimate fidelity of the steady state by avoiding measurement inefficiency and long feedback times.  However, such ``error-correction"  is only to largest order and it remains an open question whether one can further increase the measurement-less steady-state fidelity by increasing the Hilbert space, for example, or whether more conventional correction using measurement (as in typical entanglement distillation protocols) is still eventually needed.

The protocol presented  here also lends itself straightforwardly to scaling to multiple pairs of qubits. Since each qubit pair must operate at a common drive frequency to interact and entangle, this greatly avoids crosstalk between pairs operating at different frequencies.  Such a schematic is illustrated in Fig.~\ref{fig:setup} D.  Here, many pairs can operate simultaneously at their own frequency along a common transmission channel, allowing for use in more complex protocols.  One such protocol is entanglement distillation~\cite{entangdistill1996}, which would allow further improvement on the already significant steady-state concurrences achievable here, with no need to synchronize different pairs.  Thus, entangling simultaneous pairs across a common waveguide in a deterministic and persistent way could also be a boon to quantum repeater technology, where unsynchronized low-probability entanglement via postselection might run into scaling problems.

Finally, generalizing to multi-qubit entanglement using a single-excitation subspace is also possible and should straightforwardly follow the same logic. Clearly, the operator $\diss{\sum_j\kappa_j\sigma_j^-}\rho$ will have a generalized $W$ state with appropriate phase factors (summing to zero) as a dark state, and introducing near-resonant Rabi drives with the same phase factors can render the state unique and globally attractive. Studying this problem more in-depth is a direction of future research and may benefit from consideration of  ``on chip" architectures such as the atom nanophotonic waveguide interface recently employed in~\cite{gonzalez2015deterministic}.

\section{Conclusions}
\label{sec:conc}

We propose an entanglement generation and stabilisation protocol for a pair of qubits driven by single-photon transitions and interacting with a cavity mode, using only the resource of nonlocal correlated spontaneous decay from a common cavity mode or transmission line.  The qubits can either be in the same cavity or in distant, separate cavities with either cascaded or bidirectional coupling.

The technique is robust to a variety of experimental considerations and does not require coherent control (aside from constant driving of single photon transitions), measurement, collection, or feedback. The protocol has potential applicability both for short-range, high grade entanglement and long-range but finite entanglement which can be distilled via other protocols to a higher-grade concurrence.  The former is of tangible interest because global attractivity of the target state ensures that all initial and perturbed states converge to the steady-state, effectively amounting to build-in error-correction, without the need for complex pulse sequences or error-syndrome measurements and relays.  Due to its resilience towards large fractions of information loss, the protocol also enables long-distance stabilisation, which is generally very challenging as coherent control and measurement feedback loops become prohibitively costly, making environment-assisted stabilisation more advantageous.

We studied the influence of noise and loss in both uni- and bi-directional transmission channels  and conclude that significant entanglement is possible even with large losses in both cases. The solution parameters undergo a transition from symmetric to highly antisymmetric, reflecting a change of strategy from coherently driving a common steady state to catching photons that are emitted from the first cavity by the second.  While both cascaded and bidirectional coupling is feasible, the latter is highly advantageous as it suppresses the distinguishability and enhances the mitigation of the loss channels of the two qubits quadratically.  

The architectures discussed here are already within the realm of what can be realized experimentally, requiring little to no modification to existing setups \cite{roch2014observation,reiserer2014quantum, Schwartz2015}, and we have shown great robustness and potential extensibility of the techniques, rendering it a very attractive route to truly long-distance qubit entanglement.

\acknowledgments 
We thank the Kavli Institute for Theoretical Physics for hospitality and for supporting this research in part by the National Science Foundation Grant No. PHY11-25915. FM and KBW were also supported in part by the National Science Foundation under the Catalzying International Collaborations program Grant No. OISE-1158954.  EH was supported in part by the U.S. Department of Energy, Office of Science, Office of Workforce Development for Teachers and Scientists (WDTS) under the Science Undergraduate Laboratory Internship (SULI) program.  SGS acknowledges funding from the EPRSC, Ser Cymru National Research Network on Advanced Engineering and the Royal Society.  We are grateful to J. Gough, L. Martin, K. Molmer,  and M. Sarovar for fruitful and illuminating discussions.

%\bibliography{remotestabilization}

%

\end{document}